*Identification of oil starvation in hydrodynamic journal bearing using rotor vibration and Extended Kalman Filter*

Marcus Vinícius Medeiros Oliveira*

André Ricardo Fioravanti

Gregory Bregion Daniel

Corresponding author email: m263128@dac.unicamp.br

School of Mechanical Engineering, University of Campinas, Brazil

*Abstract*

Oil starvation is a critical fault in hydrodynamic bearings caused by insufficient oil supply flowrate. When late detected, this fault can deteriorate the bearings' performance and damage the rotating machine. Thus, early fault identification techniques should be applied to avoid this scenario and allow effective maintenance. However, the literature about the identification of oil starvation fault is relatively recent and very scarce. Among the few studies available, oil starvation has been identified in early stages using a purely deterministic technique, which can be jeopardized by model inaccuracy. Alternatively, this paper proposes a novel method for the identification of oil starvation faults in hydrodynamic bearings using the rotor vibrational responses and the Extended Kalman Filter (EKF), which is a stochastic-determinist state estimator that can deal with modelling and measurement errors. Computational simulations performed in this paper show that the proposed method is suitable to successfully identify the oil starvation faults in real-time even with noisy vibrational responses, thus representing a promising tool for the monitoring of rotating machines.

*1. Introduction*

Rotating machines as pumps, compressors and turbines are present in several sectors of the industry. In this kind of equipment, a key component is the bearing, whose main function is to support the rotating shaft while reducing friction. Thus, condition monitoring of hydrodynamic bearings is mandatory to assure good performance and safety of this machinery. Early fault detections are crucial to allow controlling actions or efficient maintenance programs, avoiding sudden and costly machines shutdowns.

A common type of bearing is the hydrodynamic journal bearing, which is widely used in several machines. In this configuration, the bearing surface and the rotating shaft are separated by a thin oil film that generates a hydrodynamic pressure capable to support the rotor's load. A critical fault in hydrodynamic bearings is oil starvation, which occurs when the oil supply flowrate is insufficient to fill the radial clearance in the inlet region [1]. As result, the pressure distribution is altered and, consequently, the bearing performance. Artiles e Heshmat [2] showed that reduced oil flowrates decrease the stiffness and damping of the oil film. Vincent et al. [3] verified that severe starvation condition can drastically reduce the bearing load capacity. Tanaka [4] showed that starvation increases the shaft's eccentricity inside the bearing. Based on the results obtained, these works showed that the bearing lubrication condition directly affects the rotor dynamic behavior.

The scenario of industry 4.0 has encouraged researches toward smart component capable to identify the fault occurrence. In this context, Poddar and Tandon [5,6] proposed the use of acoustic emission to detect oil starvation. However, starvation is considered in its late-stage when there is asperity interaction between bearing and shaft surfaces. Alternatively, the development of bearing failures models, including oil starvation, has allowed the application of model-based techniques to automatically quantify such faults. Generally, the strategy is to adjust the fault parameters in the model to match simulated and measured rotor vibrational responses. This approach was used by Oliveira et al. [7] to identify oil supply flowrates. However, it is a purely deterministic approach, strongly dependent on model accuracy. An alternative technique that has not been explored in oil starvation identification is the use of state-estimators, which can infer internal parameters, as faults, from inputs and outputs of a system. Among these techniques, the Kalman Filter is probably the most popular due to its optimality and capability to deal with model and measurement uncertainties.

The Kalman Filter (KF) is a recursive state-estimator that combines model predictions and system measurements, both assumed with stochastic error for a better estimate of a measured or unknown state of linear dynamic systems [8]. When the systems are nonlinear, the Extended Kalman Filter (EKF) [9] or the Unscented Kalman filter (UKF) [10] are applied instead. Typical uses of Kalman filtering include smoothing noisy data and providing estimates of parameters of interest [11]. As a parameter estimator, the filter has been applied for fault identification in several applications, such as sensors [12-14], actuators [15], and chemical process [16].

In rotordynamics, there are some works in which Kalman filtering is used to identify different parameters and faults. Fritzen and Seibold [17] successfully applied the EKF to identify the damping factor of a rotor with a cracked shaft, as well as the crack depth. The authors were also able to identify the dynamic coefficients of a seal on a turbine pump. Provasi et al. [18] used the EKF to identify the modal parameters of supporting structures of rotating machines. Cao et al. [19] applied the UKF to identify the gearbox faults in wind turbines from power measurements. Michalski and Souza [20] used the KF with vibration measurements to identify the unbalance in a Laval rotor and in an industrial turbocharger. With a similar method, Zou et al. [21] analyzed the influence of the sensor position and the measurement noise in the unbalance identification.

Regarding the bearings, there are few applications of EKF and KF to predict rolling bearing degradation [22,23]. The literature about parameter identification using KF is even more reduced for hydrodynamic bearings, despite its wide application in industry. Miller and Howard [24] used the Extended Kalman filter to identify the dynamic coefficients of a gas-foil bearing. It is also worth mentioning the identification of bearing coefficient using another recursive filtering technique named Recursive Least Squares (RLS), as proposed by Coraça and Junior [25]. In these works, the identification is related to generalized parameters that can indicate the eventual decreasing of the bearing performance but do not allow to directly quantify the physical reason.

In the case of hydrodynamic journal bearing, there is no literature about directly detecting physical faults through state-estimators. Thus, the current work proposes a method to identify the bearings oil supply flowrate from the rotor vibrational response using the Kalman filter, allowing to monitor the hydrodynamic conditions and identify the oil starvation faults. For this, the rotor-bearings system is modeled in a way that the oil supply flowrate can be directly accessed as a state variable. Thus, the model is incorporated into an Extended Kalman filter algorithm, in which the oil flowrates are estimated while the predicted rotor displacements are corrected with measured rotor vibration. The proposed method is successfully applied to a turbine model under different lubrication conditions, in which the numerical simulations were performed to identify oil starvation due to leakage or clogging in the oil line distribution. The results obtained in this work show that EKF is promising for the identification of oil starvation faults in hydrodynamic bearings, which contributes to the improvement and development of new tools for condition monitoring and fault diagnosis in rotating machines.

## 2. Methodology

In this section, the numerical modelling of the rotor-bearing system and the EKF formulation are presented. First, the bearing model is presented in section 2.1. Next, the entire rotating system model and its adaptation to oil supply flowrate estimation are presented in sections 2.2 and 2.3, respectively. Finally, the EKF formulation is presented in section 2.4.

### 2.1 Bearing model

The main parameters of the hydrodynamic journal bearing are shown in Fig. 1, being $e_X$ and $e_Y$ the shaft eccentricity in horizontal and vertical directions, R the shaft radius, $\Omega$ the shaft rotational speed, $x$ the circumferential direction and $h$ the oil film thickness.

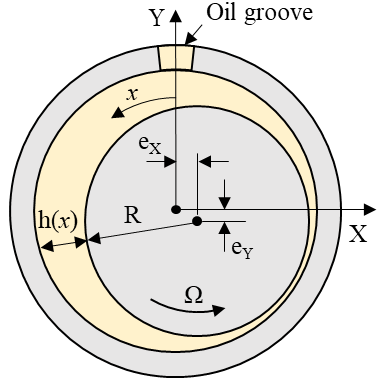

Fig. 1 – Journal bearing model.

In the present work, a mass-conservation model is used to solve the hydrodynamic bearing by Finite Volume Method (FVM), in which the oil supply flowrate is considered as an input variable [7]. In this approach, the pressure distribution in the oil film is described by a modified Reynolds Equation, namely the $p - \theta$ model:

$$\frac{\partial}{\partial x}\left(\frac{h^3}{12\mu}\frac{\partial p}{\partial x}\right) + \frac{\partial}{\partial z}\left(\frac{h^3}{12\mu}\frac{\partial p}{\partial z}\right) = \frac{\partial}{\partial x}\left(\frac{U}{2}h\theta\right) + \frac{\partial(h\theta)}{\partial t} \tag{1}$$

where $p$ is the hydrodynamic pressure, $\theta$ is the fluid fraction, $z$ is the axial direction, $\mu$ is the oil dynamic viscosity, $U$ is the linear velocity on the shaft surface and $t$ is the time.

Using the FVM, the oil film is discretized and the Eq.1 is integrated in the circumferential and axial directions of the bearing, performing a flow balance through the finite volumes in the bearing domain. Thus, the oil supply flowrate is added to the flow balance of the volumes at the bearing groove region. This integration results in equations of $p$ and $\theta$ for every finite volume, which are dependent on its neighbouring volumes. From initial distributions, the $p$ and $\theta$ distributions are obtained with an iterative procedure in which the cavitation condition is assured [26]. Once converged, the pressure distribution is integrated along the bearing domain to obtain the hydrodynamic forces acting on the shaft.

To reduce computational time for numerical simulations, the hydrodynamic forces are linearized around the equilibrium position, as proposed by Lund [27]:

$$F_{h_X} = K_{XX}(e_X - e_{X0}) + K_{XY}(e_Y - e_{Y0}) + C_{XX}\dot{e}_X + C_{XY}\dot{e}_Y \tag{2}$$

$$F_{h_Y} = K_{YY}(e_Y - e_{Y0}) + K_{YX}(e_X - e_{X0}) + C_{YY}\dot{e}_Y + C_{YX}\dot{e}_X \tag{3}$$

where $K_{ij}$ are the derivatives of the hydrodynamic forces in relation to the shaft's displacement, representing the equivalent stiffness coefficients, and $C_{ij}$ are the derivatives of the hydrodynamic forces in relation to the shaft's linear velocities, representing the equivalent damping coefficients.

### 2.2 Rotating system model

The Finite Element Method (FEM) is used to model a typical rotating system composed of a flexible shaft, rigid discs and two hydrodynamic bearings. The rotor is discretized by Timoshenko beam elements with four degrees of freedom (dof) per node, being two linear displacements $V$ and $W$ (in $X$-direction and $Y$-direction, respectively) and two angular displacements B and $\Gamma$ (around axes $X$ and $Y$, respectively), as shown in Fig. 2.

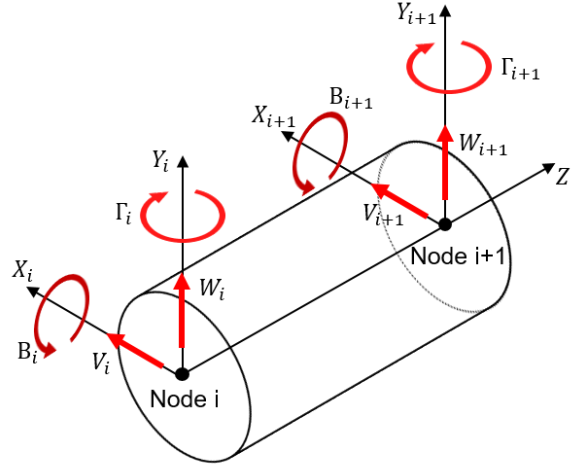

Fig.2 – Timoshenko beam element.

Assembling the equations of each element, the global equation of motion of all nodes is obtained as:

$$\boldsymbol{M}\ddot{\boldsymbol{r}}(t) + (\boldsymbol{C} + \Omega\boldsymbol{G})\dot{\boldsymbol{r}}(t) + \boldsymbol{K}\boldsymbol{r}(t) = \boldsymbol{u}(t) \tag{4}$$

where $\boldsymbol{M}$, $\boldsymbol{C}$, $\boldsymbol{G}$ and $\boldsymbol{K}$ are the inertia, damping, gyroscopic and stiffness matrices of the rotor, $\Omega$ is the rotor's rotational speed, $\boldsymbol{r}$ is the displacements vector and $\boldsymbol{u}$ is the external forces vector, including the rotor weight.

The rigid discs are modeled by adding their matrices of inertia and gyroscopic inside the matrices $\boldsymbol{M}$ and $\boldsymbol{G}$, in the degrees of freedom corresponding to the rotor's nodes coupled to the discs. In addition, the unbalance forces $F_{u_X}$ and $F_{u_Y}$ are included in the discs nodes inside the vector $\boldsymbol{u}$, in order to represent the unbalance excitation:

$$F_{u_X} = m_\varepsilon\Omega^2\cos(\Omega t + \phi) \tag{5}$$

$$F_{u_Y} = m_\varepsilon\Omega^2\sin(\Omega t + \phi) \tag{6}$$

being $m_\varepsilon$ and $\phi$ the unbalance moment and phase.

Finally, the hydrodynamic forces $F_{h_X}$ and $F_{h_Y}$ are linearized by equivalent coefficients of damping and stiffness, as described in the previous section. These forces are included inside the vector $\boldsymbol{u}$, acting on the rotor's nodes at the position of the bearings.

### 2.3 Augmented state-space including the oil supply flowrates

As a state observer, the KF is capable to estimate the internal state vector of a linear system from known inputs (forces acting on the system) and measured outputs. Thus, Eq. 4 is written in a state-space representation as:

$$\dot{\boldsymbol{s}}(t) = \boldsymbol{A}.\boldsymbol{s}(t) + \boldsymbol{B}.\boldsymbol{u}(t) \tag{7}$$

$$\boldsymbol{z}(t) = \boldsymbol{H}.\boldsymbol{s}(t) \tag{8}$$

where $\boldsymbol{s} = [\boldsymbol{r}^T \ \dot{\boldsymbol{r}}^T]^T$ is the state vector, $\boldsymbol{z}$ is the output vector, $\boldsymbol{H}$ is the output matrix, $\boldsymbol{A}$ is the state matrix and $\boldsymbol{B}$ is the input matrix, being defined as:

$$\boldsymbol{A} = \begin{bmatrix} \boldsymbol{0} & \boldsymbol{I} \\ -\boldsymbol{M}^{-1}\boldsymbol{K} & -\boldsymbol{M}^{-1}(\boldsymbol{C} + \Omega\boldsymbol{G}) \end{bmatrix} \tag{9}$$

$$B = \begin{bmatrix} 0 \\ M^{-1} \end{bmatrix} \tag{10}$$

Once the filtering process is applied to sampled data, the continuous-time state-space model is discretized into discrete-time domain:

$$s_{k+1} = A_d . s_k + B_d . u_{k+1} \tag{11}$$

$$z_k = H . s_k \tag{12}$$

where $k$ is the current discrete time-step. The matrices $A_d$ and $B_d$ are the discretized state and input matrices, which can be obtained using the following relation with the matrix exponential:

$$e^{\begin{bmatrix} A & B \\ 0 & 0 \end{bmatrix} \Delta t} = \begin{bmatrix} A_d & B_d \\ 0 & I \end{bmatrix} \tag{13}$$

where $\Delta t$ is the sampling period.

To use the KF to estimate the oil supply flowrates $q = [q_1 \ q_2]^T$, Eq. 11 and 12 should be rewritten including the oil supply flowrates in the state vector.

$$s_{a_{k+1}} = \begin{bmatrix} A_d & 0 \\ 0 & 1 \end{bmatrix} . s_{a_k} + \begin{bmatrix} B_d \\ 0 \end{bmatrix} . u_{k+1} \tag{14}$$

$$z_k = [H \quad 0] . s_{a_k} = H_a . s_{a_k} \tag{15}$$

where $s_a = [r^T \ \dot{r}^T \ q^T]^T$ is the augmented state vector containing the displacements, velocities and oil supply flowrates.

Consequently, the augmented state vector becomes coupled to the hydrodynamic forces of the vector $u$, since the equivalent coefficients of stiffness and damping depend on the oil supply flowrates. Thus, the system represented by Eq. 14 is no longer linear and will be summarized as a general nonlinear system:

$$s_{a_{k+1}} = f(s_{a_k}, u_{k+1}) \tag{16}$$

### 2.4 Identification of the oil supply flowrate using EKF

Considering the existence of error in the modeling and noise in the measurements, one assumes that the true state and output of the rotating system evolves according to:

$$s_{a_{k+1}} = f(s_{a_k}, u_{k+1}) + w_k \tag{17}$$

$$z_k = H_a . s_{a_k} + v_k \tag{18}$$

where $w_k$ and $v_k$ are process and measurement noise drawn from zero-mean multivariate normal distributions with covariance $Q$ and $R$, respectively. Once defined the state function of the rotating system and its covariance matrices, the Kalman filtering process is applied to estimate the states based on model predictions and measurements. The state estimates are noted as $\hat{s}_a$ to be distinguished from the unknown true state $s_a$.

Since the state function is nonlinear, it is necessary to apply the EKF, in which the state function is linearized with a Taylor approximation at the previous estimated state [9]. Thus, from an initial estimated state $\hat{s}_{a_{k|k}}$ with error covariance $\hat{P}_{k|k}$, an a-priori estimate of the next state $\hat{s}_{a_{k+1|k}}$ and its error covariance $\hat{P}_{k+1|k}$ can be predicted as following:

$$\hat{s}_{a_{k+1|k}} = f(\hat{s}_{a_{k|k}}, u_{k+1}) \tag{19}$$

$$\hat{P}_{k+1|k} = J_{k+1} . \hat{P}_{k|k} . J_{k+1}^T + Q \tag{20}$$

where $J_{k+1}$ is the Jacobian matrix:

$$J_{k+1} = \frac{\partial f}{\partial s}\Big|_{\hat{s}_{a_{k|k}}, u_{k+1}} \tag{21}$$

Then, the Kalman filter gain $K_{k+1}$ is computed and the a-priori estimate and its error covariance are updated with measurements $z_{k+1}$, resulting in an a-posteriori state estimate $\hat{s}_{k+1|k+1}$ with error covariance $\hat{P}_{k+1|k+1}$:

$$K_{k+1} = \hat{P}_{k+1|k}H_a{}^T[H_a\hat{P}_{k+1|k}H_a{}^T + R]^{-1} \tag{22}$$

$$\hat{s}_{a_{k+1|k+1}} = \hat{s}_{a_{k+1|k}} + K_{k+1}[z_{k+1} - H_a\hat{s}_{a_{k+1|k}}] \tag{23}$$

$$\hat{P}_{k+1|k+1} = [I - K_{k+1}H_a]\hat{P}_{k+1|k} \tag{24}$$

For every time step $k$, the value of the oil supply flowrates $q$ will change with the state estimate and the bearing coefficients may be recomputed. To reduce the computational cost, the coefficients will be interpolated from values previously computed for certain oil supply flowrates. For identification of supposing constant unknown oil supply flowrates, the procedure ends when $q$ converges. Furthermore, the filter can be kept running as long as the outputs in the rotating system can be measured to detect real-time variation of the oil supply flowrate in the bearing. The identification procedure is summarized in Fig. 3.

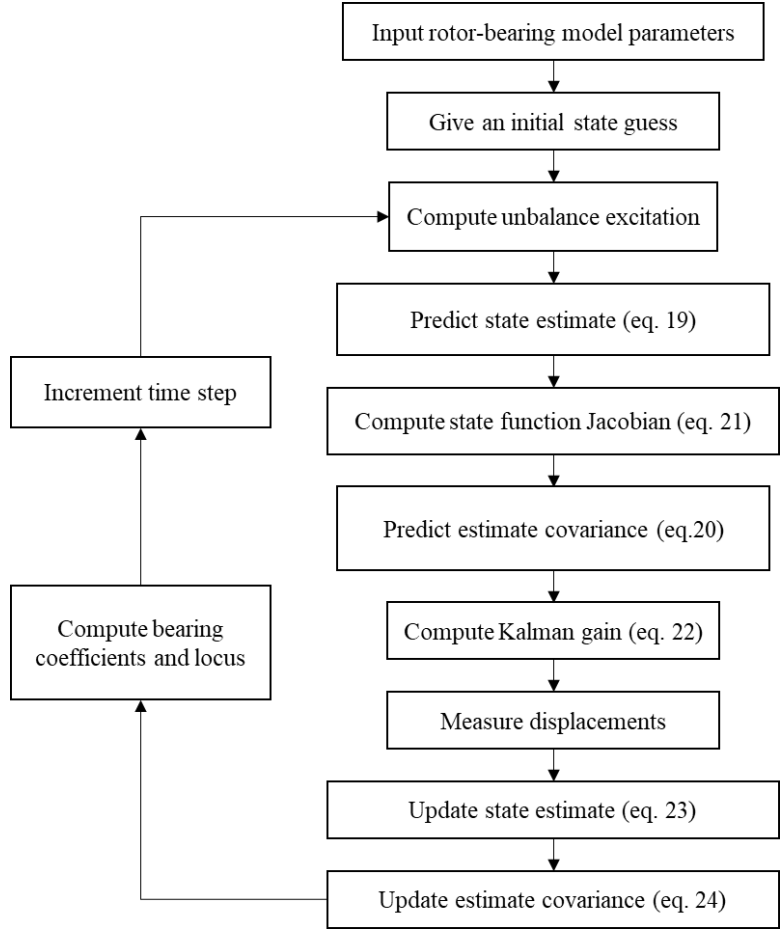

Fig. 3 – Identification procedure of the oil supply flowrate using EKF

## 3. Numerical results

This section presents the numerical results obtained with the proposed method, considering its application in a rotating system supported by hydrodynamic journal bearings under different lubrication conditions. The model of the rotor used in the computational simulations represents a generic turbine, as shown in Fig. 4. This rotor is composed of 20 beam elements, three rigid discs and two identical cylindrical bearings, whose characteristics are detailed in Tables 1, 2 and 3. The nominal rotational speed of the rotor is 75 Hz, lying between the first critical speed (~55 Hz) and the instability threshold (~95 Hz). The discs and shaft are composed of steel with Young Modulus $E = 210\,GPa$, Poisson ratio $\nu = 0.3$ and density $\rho = 7850\,kg/m^3$. An unbalance force is considered acting on the central disc (node 13), the unbalance moment is defined based on the class G2.5 for gas and steam turbines [28]. The total rotor weight ($W_Y = 6545.6\,N$) is equally distributed at the bearing 1 (node 6) and bearing 2 (node 20), resulting in a loading of $3272.8\,N$ in each bearing.

Inlet pressure of 0.5 bar (50 kPa) was considered at the bearing groove to define a nominal lubrication condition. At this condition, the nominal oil supply flowrate through the bearings is 596.3 ml/min. From this value, different lubrication conditions were simulated to test the proposed identification method. The stiffness and damping coefficients of the hydrodynamic bearing are interpolated from values pre-computed for 7 oil supply flowrates between 50% and 150% of the flooded lubrication threshold $q_t = 546\,ml/min$.

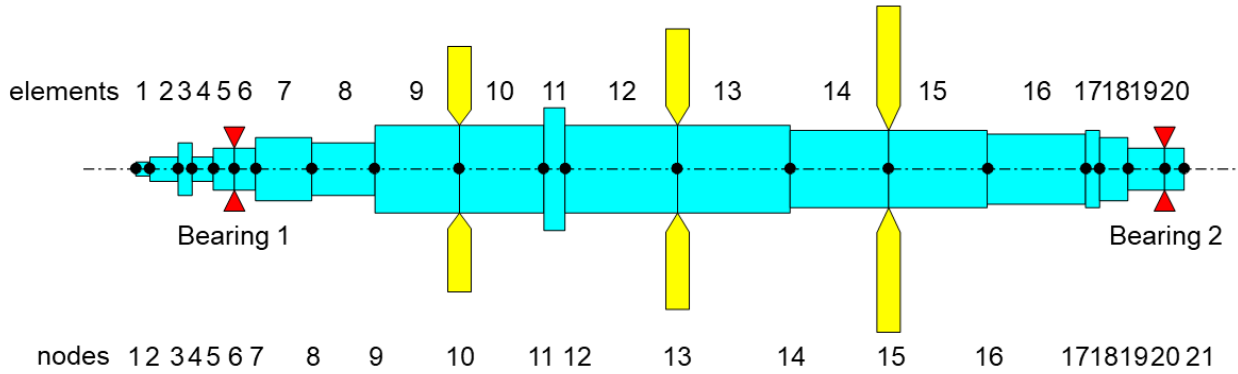

Fig. 4 - Model of the generic turbine used in the computational simulations.

Table 1
Shaft element dimensions.

| Element Number | Length [mm] | Diameter [mm] |
|---|---|---|
| 1 | 30.0 | 30.0 |
| 2 | 60.0 | 52.5 |
| 3 | 30.0 | 112.5 |
| 4 | 45.0 | 52.5 |
| 5,6 | 45.0 | 90.0 |
| 7 | 120.0 | 135.0 |
| 8 | 135.0 | 112.5 |
| 9,10 | 180.0 | 187.5 |
| 11 | 45.0 | 262.5 |
| 12,13 | 240.0 | 187.5 |
| 14,15 | 210.0 | 165.0 |
| 16 | 210.0 | 150.0 |
| 17 | 30.0 | 165.0 |
| 18 | 60.0 | 135.0 |
| 19 | 78.4 | 90.0 |
| 20 | 41.6 | 90.0 |

Table 2
Disc element dimensions.

| Disc Node | Width [mm] | External Diameter [mm] |
|-----------|------------|------------------------|
| 10 | 50.0 | 525.0 |
| 13 | 50.0 | 600.0 |
| 15 | 50.0 | 697.5 |

Table 3
Bearing and lubricant oil properties.

| Parameters and units | Value |
|----------------------|-------|
| Radius [mm] | 45.0 |
| Width [mm] | 70.0 |
| Radial clearance [μm] | 120.0 |
| Oil dynamic viscosity [Pa.s] | 0.094 |
| Groove angular position [°] | 0 |
| Groove circumferential length [mm] | 16.2 |
| Groove width [mm] | 35.0 |

For each lubrication condition, the rotor vibrational response was simulated to be used as measured data in the EKF. This "measured" response was obtained from the rotating system model, integrated over time by the *ode23t* solver of the MATLAB with a fixed time step of 1ms (sampling time). The measurements considered were the shaft's displacements inside the bearings 1 and 2 (nodes 6 and 20) in the horizontal and vertical directions. Firstly, the sensitivity of the displacements in relation to the oil supply flowrate was analyzed. Then, the displacements were contaminated with zero-mean uncorrelated white noise and different identification tests were performed.

### 3.1 Sensitivity of the rotor vibrational response in relation to oil supply flowrate

Since the parameter identification using EKF is based on state observations, the displacements must be distinguishable for different values of oil supply flowrates. Fig. 5 shows the shaft's orbits inside the bearing 1 and 2, while $q1$ and $q2$ decrease together from 100% to 75% of the nominal oil supply flowrate. It can be seen that the orbits amplitudes decrease and the eccentricity increases, as the oil supply flowrates decrease. In addition, the orbits became closer to each other near the nominal flowrate, for both bearings. However, the amplitude of the displacements inside the bearing 2 is higher than those inside the bearing 1.

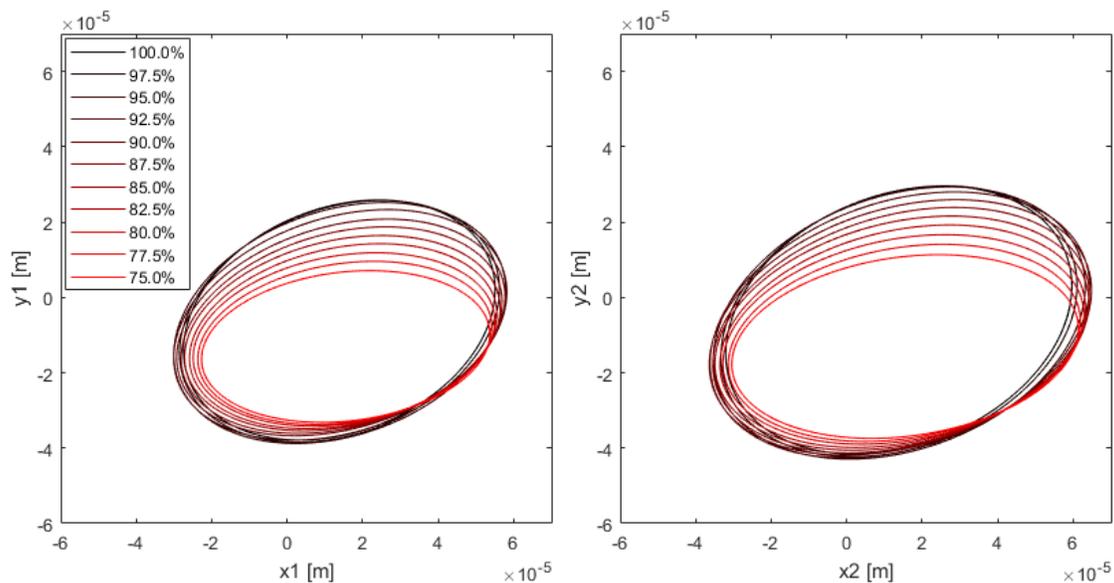

Fig. 5 – Orbits variation during the oil supply flowrate decrease.

Despite the equally distributed static load, the rotor geometry and the unbalance excitation are asymmetric, causing the observed behavior. To better visualize the orbits variation, the eccentricity and the variation of the amplitudes were quantified for different levels of oil supply flowrate (Table 4). As the oil supply flowrate approaches the nominal value, the eccentricity variation becomes lower, making difficult the distinction of the orbits. In addition, the horizontal amplitude variation is reduced between 95% and 90% of nominal flowrate, while the vertical amplitude variation is reduced near 100%, mainly in the bearing 1. These effects tend to increase the error of the identified flowrates when both bearings are in a flooded condition. Furthermore, the lower amplitudes and variation inside the bearing 1 may make it difficult to distinguish the lubrication conditions in presence of measurement noise, probably reducing the accuracy of the identification of the oil supply flowrate (q1) in this bearing.

Table 4
Orbits variation for q1 and q2 decreasing.

| Variation Interval of the Oil Supply Flowrate (%) | 100-95 | 95-90 | 90-85 | 85-80 | 80-75 |
|---|---|---|---|---|---|
| Eccentricity variation bearing 1 (µm) | 0.78 | 1.22 | 1.55 | 1.74 | 1.92 |
| Eccentricity variation bearing 1 (µm) | 0.78 | 1.22 | 1.55 | 1.74 | 1.92 |
| Amplitude X variation bearing 1 (µm) | -5.39 | 0.95 | 3.37 | 3.71 | 3.54 |
| Amplitude X variation bearing 2 (µm) | -6.84 | -2.72 | 2.19 | 3.42 | 3.79 |
| Amplitude Y variation bearing 1 (µm) | -0.22 | 6.44 | 5.72 | 5.77 | 5.73 |
| Amplitude Y variation bearing 2 (µm) | -1.10 | 4.42 | 5.70 | 6.50 | 6.87 |

The orbit sensitivity was also analyzed for different variations of oil supply flowrate in the bearings 1 and 2. According to Table 5, a reduction of the oil supply flowrate only in the bearing 1 (q1) causes an orbit variation in both bearings, evidencing the dependency between the dynamic of the bearings. Although the orbit changes inside the bearing 2, its eccentricity remains the same since the oil supply flowrate is fixed. This sensitivity analysis has shown that the displacements are related to the oil flowrates, thus it is expected that the EKF will be able to correctly adjust the flowrates values while observing the measured vibration.

Table 5
Orbits variation for q2 nominal and q1 decreasing.

| Variation Interval of the Oil Supply Flowrate (%) | 100-95 | 95-90 | 90-85 | 85-80 | 80-75 |
|---|---|---|---|---|---|
| Eccentricity variation in bearing 1 (µm) | 0.78 | 1.22 | 1.55 | 1.74 | 1.92 |
| Eccentricity variation in bearing 2 (µm) | 0.00 | 0.00 | 0.00 | 0.00 | 0.00 |
| Amplitude variation X bearing 1 (µm) | -13.83 | -13.27 | 0.57 | 2.84 | 3.32 |
| Amplitude variation X bearing 2 (µm) | 7.30 | 10.36 | 0.74 | -1.32 | -2.42 |
| Amplitude variation Y bearing 1 (µm) | -5.95 | -1.38 | 6.13 | 7.64 | 8.12 |
| Amplitude variation Y bearing 2 (µm) | 5.02 | 6.20 | -0.40 | -1.71 | -2.34 |

*3.2 Identification of the oil supply flowrate*

The identification was tested for two different levels of noise contaminating the measured displacements, being one with standard deviation of $\sigma_v = 1\ \mu m$ and the other with $\sigma_v = 2\ \mu m$, as shown in Fig. 6.

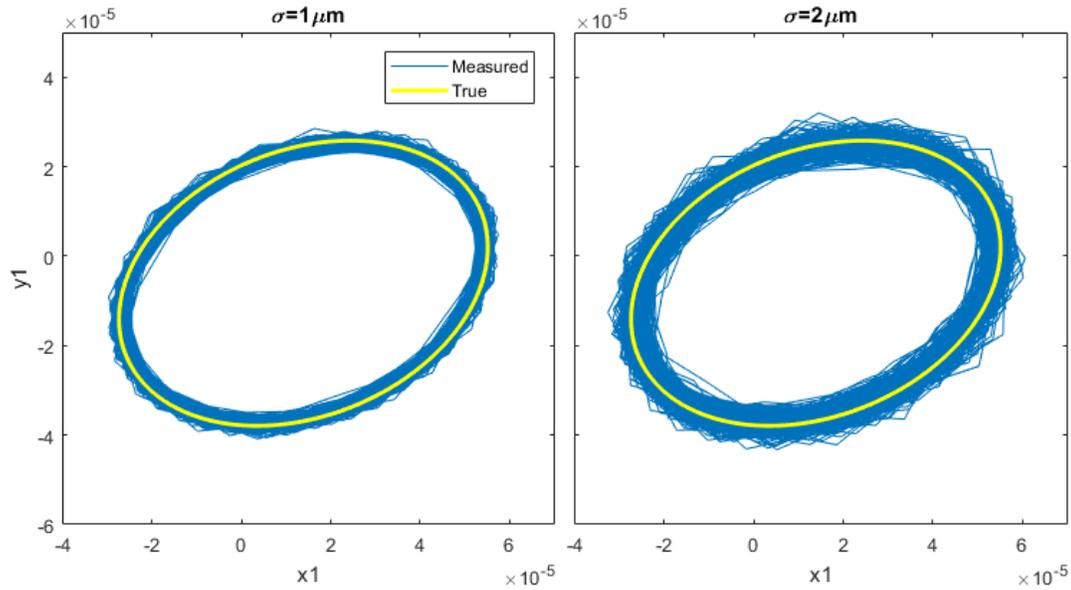

Fig. 6 – Noisy displacement measurements in the bearing 1 (node 6).

In addition, the shaft's velocities in the horizontal and vertical directions were numerically obtained from the shaft's displacements, augmenting the information about the real system without extra sensors. Thus, the measurement covariance matrix R was defined as an 8x8 diagonal matrix whose values are $\sigma_v{}^2$. Different variances for the process noise in the states were tested until achieving convergence and reasonable error in the estimate of the oil supply flowrates. For the application in this work, a suitable covariance matrix was obtained considering a standard deviation of 0.1 ml/min for the noise of the oil supply flowrate and a standard deviation of 10 μm and 10 μm/s for the noises of the displacement and velocities, respectively. Finally, the initial estimate of the displacements and velocities were set to zero (null initial conditions), while the initial estimate of the oil supply flowrates was chosen to be the nominal value of 596.3 ml/min.

### 3.2.1 Identification of constant oil supply flowrate

Firstly, the identification method was performed for different lubrication conditions considering the oil supply flowrates as constant in both bearings. Thus, it was considered combinations of oil supply flowrates in each bearing varying from 100% to 75% of the nominal value (596.3 ml/min), covering both flooded and starved lubrication conditions. For each case, the measured displacement was generated for a period of 10s, but the data before 5s was discarded to avoid numerical transient. The identified value for the oil supply flowrates was assumed as the average value estimated after the convergence. Hence, the relative error between the identified and true values of the oil supply flowrates can be computed for all cases.

For the lower measurement noise ($\sigma_v = 1.0\ \mu m$), the computed error for the identified oil supply flowrate in the bearings 1 and 2, i.e. q1 and q2, are shown in **Erreur ! Source du renvoi introuvable.**Fig. 7. The oil supply flowrates are satisfactorily identified in both bearings, with a maximum relative error of 5.06%, average relative error of 1.82% and standard deviation of 0.99%. The identification on bearing 1 presented a higher average relative error of 2.28%, against 1.37% observed in bearing 2, as suggested by the results in section 3.1. There is not a clear pattern in the error distribution, but an asymmetric and nonlinear relation with the oil supply flowrate, which can be attributed to the nonlinear behavior of the rotating system in relation to the oil supply flowrate.

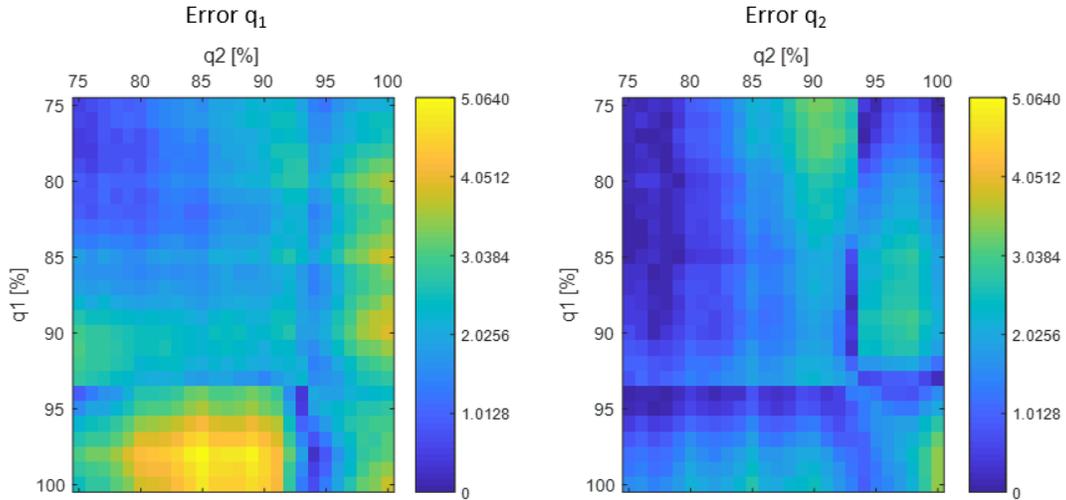

Fig. 7 – Relative error for the identified oil supply flowrates with low noisy data ($\sigma_v = 1.0\ \mu m$).

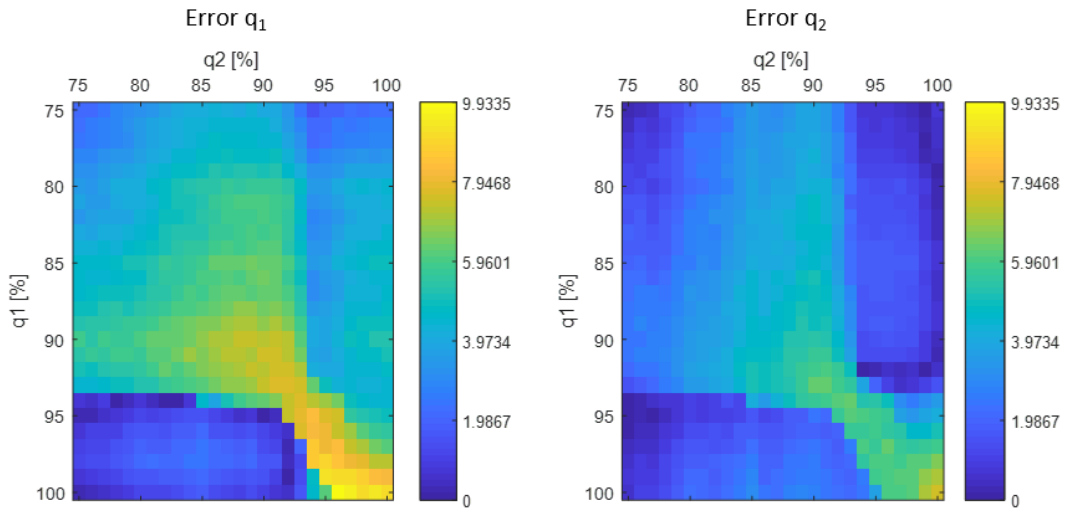

Fig. 8 – Relative error for the identified oil supply flowrates with high noisy data ($\sigma_v = 2.0\ \mu m$).

Fig. 8 shows the relative errors of the identified oil supply flowrates considering the higher measurement noise ($\sigma_v = 2\ \mu m$). In this case, the relative errors are higher for identified flowrates close to the nominal value in both bearings simultaneously. As expected, the higher noise level increases the relative errors, resulting in a maximum value of 9.93%. Despite that, the average relative error and the standard deviation remain satisfactory, whose values are 3.54% and 1.97%, respectively. Finally, the observed results are summarized in Table 6.

Table 6
Average and maximum errors in the oil flowrate identification

| Measurement noise standard deviation ($\mu m$) | Identified oil flowrate average error (%) | Standard deviation (%) | Maximum error (%) |
|---|---|---|---|
| 1.0 | 1.82 | 0.99 | 5.06 |
| 2.0 | 3.54 | 1.97 | 9.93 |

Observing the low relative errors obtained for the several simulations, it is possible to conclude that the EKF is able to identify the oil supply flowrates in the hydrodynamic bearings with satisfactory flexibility and accuracy. In addition, the proposed method seems promising and efficient, since the identification requires only a few seconds of data acquisition, as shown in Fig.9 for the case of the highest error in both measurement noise.

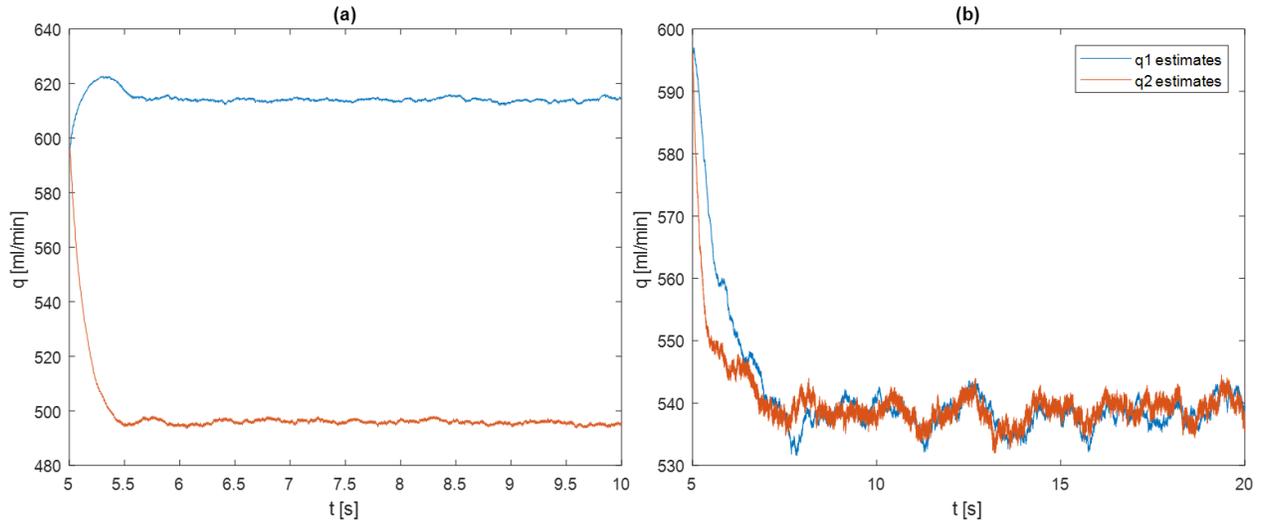

Fig. 9 – Evolution of the estimates q1 and q2 for the cases with highest relative error. a) q1=98% (584.3) q2=85% (506.8) and σ=1.0 μm, b) q1=100% (596.3) q2=96% (572.45) and σ=2.0 μm.

*3.2.2 Identification of the real-time variation of oil supply flowrates*

In this section, the EKF was tested to identify variations of the oil supply flowrate in real-time, simulating different disturbances in the oil distribution lines. Firstly, it was simulated that the oil supply flowrates in the bearings 1 and 2 change equally, dropping from 100% to 75% of the nominal value (i.e. from 596.3 to 447.2 ml/min), in order to represent a generalized flowrate drop of 25%. In practical situations, this problem could indicate a fault in the oil pump or leakage/obstruction of the main oil distribution line. The oil supply flowrate drop was simulated using a sigmoid transition between the initial and final flowrate value. Three scenarios were tested for the simulation of this flowrate drop, namely, a slow drop, an intermediate drop and a sudden drop. The experimental signal was generated for 20s with the oil supply flowrate drop occurring around 10s. Again, the identification was tested with two levels of measurements noises, being σ=1.0 μm and σ=2.0 μm.

Fig. 10 shows the evolution of estimated oil supply flowrate for the three kind of drops considering the lower level of measurements noise (σ=1.0 μm). It is possible to note that the EKF successfully identifies the decrease of oil supply flowrate, with a good agreement during all transitions.

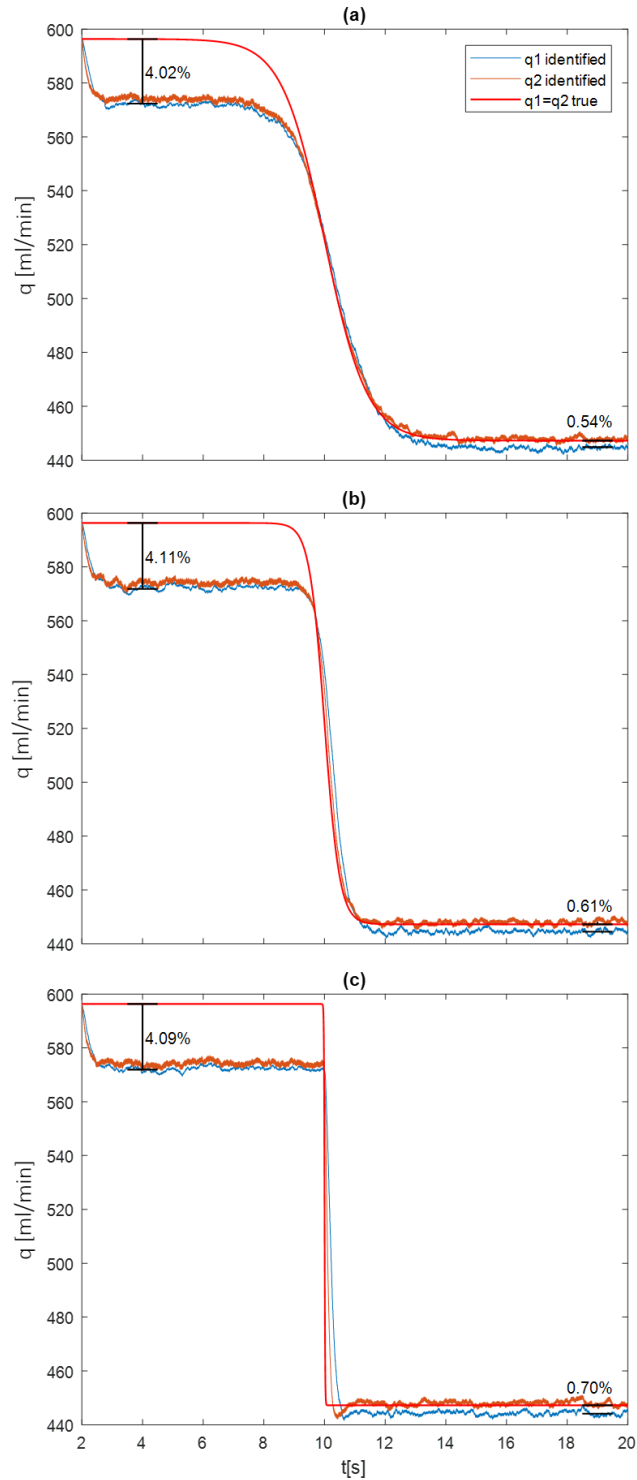

Fig. 10 – Identification of the oil supply flowrate in real-time: a) slow drop b) intermediate drop and c) sudden drop.

According to Fig. 10, the sensitivity of both bearings is practically the same under extremal oil supply flowrates, i.e. 100% and 75%. Thus, it occurs only errors before and after the oil flowrate drop. For all drops simulated, the relative errors of the identified flowrates before the drop are around 4%, while the relative errors of identified flowrates after the drop are less than 1%. This behavior is in agreement with the previous section, since the identification of low oil supply flowrates is more accurate than that of nominal flowrates. Considering the higher measurements noise (σ=2.0 μm), the errors increase as expected, being around 9% before the drop and 2% after that. These results show that the EKF is capable to rapidly detect a real-time abrupt deterioration in the hydrodynamic bearings lubrication condition, which are important for monitoring the rotating machine health.

Considering that the oil supply flowrates of the bearings 1 and 2 can be independently damaged, the identification was further applied to situations where the oil flowrate variations are not the same for both bearing. Fig. 11 shows the evolution of estimated oil supply flowrates assuming an intermediate drop of 25% only in one of bearings, while the other remains the nominal flowrate. Fig. 11a shows the flowrate drop occurred only in bearing 1 and Fig. 11b shows the flowrate drop occurred only in bearing 2. In practical situations, this condition could indicate a leakage in the distribution line derived to one of the bearings.

Again, the EKF successfully identifies the initial nominal flowrates and the oil flowrate drop in the correct bearing for both cases. The errors are similar to those observed previously: around 4% for the nominal flowrate and less than 1% for the low flowrate. However, the errors for nominal oil flowrates are also reduced after the drop, becoming about 2%. For the higher measurements noise (σ=2.0 μm), the errors of the identified initial nominal flowrates increase to 9%, but, after the drop, the remaining nominal flowrate and the low flowrate are successfully identified with an error of 1% and 2% respectively.

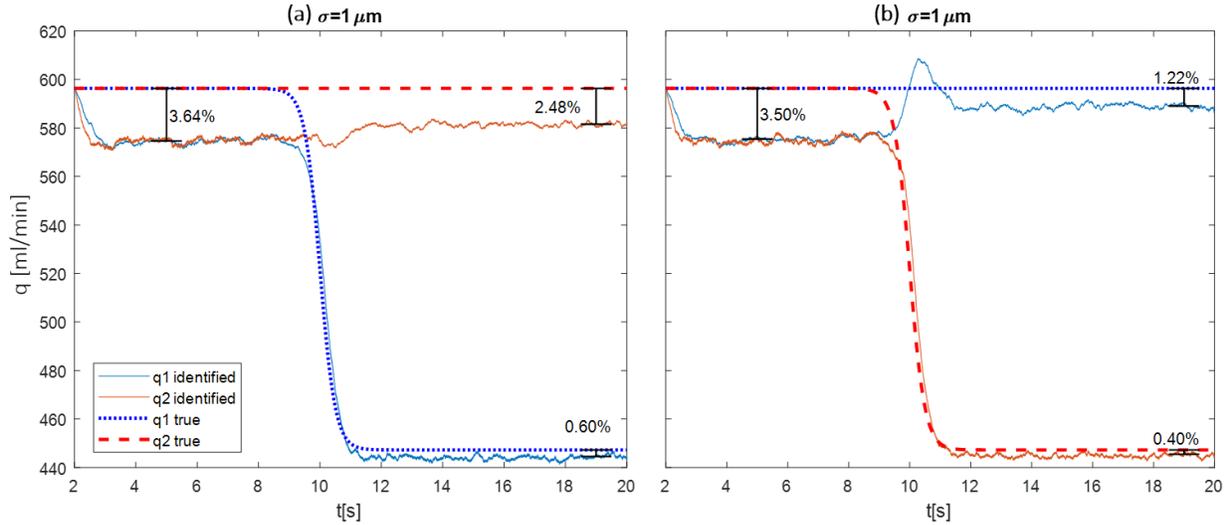

Fig. 11 Identification of the oil supply flowrate in real-time: a) drop only in the bearing 1. b) drop only in the bearing 2.

Another possible scenario was tested considering that the oil line of one bearing is partially clogged, causing a flowrate drop of 25% in the bearing of this line, while a flowrate increase of the same amount occurs in the bearing of other lines, i.e. changing from 596.3 to 745.4 ml/min. Fig 12a shows the identification for the situation with flowrate drop in the bearing 1 and Fig 12b shows the identification with flowrate drop in the bearing 2. Again, the EKF is capable to correctly identify the initial and final oil flowrates in both bearings. The relative errors are still the same for the initial flowrate (nominal value) and final low flowrate, being around 4% and 1% respectively. However, the relative error increases for the final high flowrates, being around 12% in the bearing 1 and 9% in the bearing 2. Although the increase is perceptible and identified in the correct bearing, the EKF is not able to accurately identify the oil supply flowrate due to the low sensitivity of the rotor vibrational response under flooded conditions in the bearings, as previously shown in section 1. For cases with higher measurements noise, the relative error of the identified initial flowrate increases to 9% and the relative error of the identified final flowrate after the drop remains around 1%. However, the relative error of the identified final flowrate after the increase reaches 14%. This behavior was expected because it represents a critical case with a high level of measurement noise and flooded lubrication condition in the bearing.

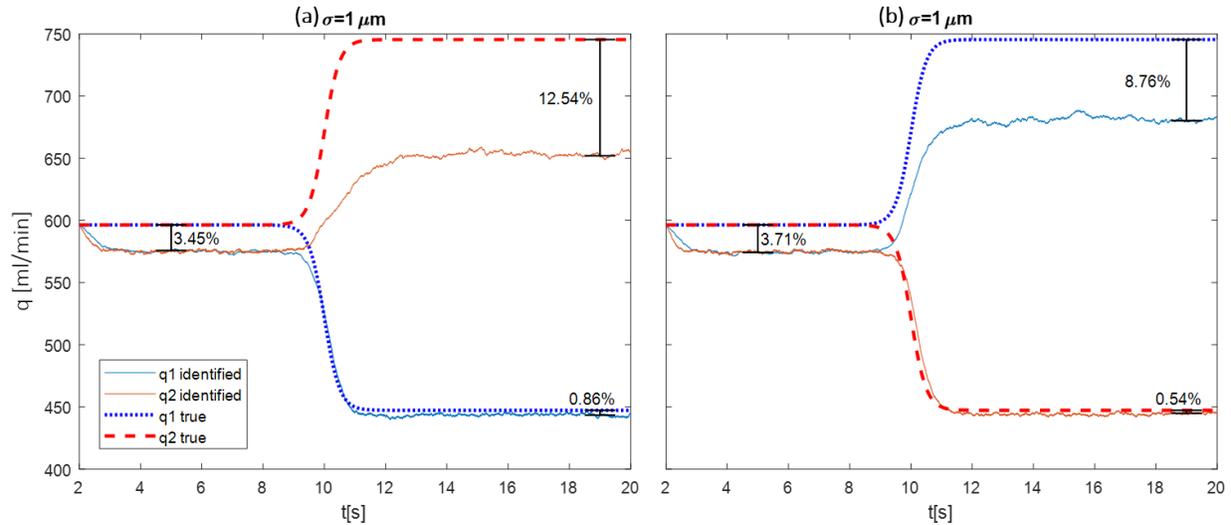

Fig. 12 Identification of the oil supply flowrate in real-time: a) drop in the bearing 1 and increase in the bearing 2, b) drop in the bearing 2 and increase in the bearing 1.

The numerical simulations performed in this research showed that the oil supply flowrates can be estimated from the rotor vibrational response using the EKF, thus allowing to successfully identify oil starvation faults in hydrodynamic journal bearing. It is important to highlight that the oil starvation fault identification and real-time monitoring is the main focus of this study since it can cause a dangerous condition during the operation of rotating machinery.

## 4. Conclusions

This paper proposes a novel method to estimate the oil supply flowrate in hydrodynamic journal bearings using only the rotor's vibrational measurement and the Extended Kalman Filter (EKF) algorithm. For this, a numerical approach was developed to access the bearings oil flowrates as states of the state-space model of the dynamic system.

Based on the computational simulations of a generic turbine, the results showed that the rotor vibrational responses in the time domain can be distinguished for different lubrication conditions, being the dynamic behavior of the rotor significantly sensitive to variation of oil supply flowrate under oil starvation condition. Thus, the numerical simulations performed in this research showed that the proposed method is effective to identify faults of oil starvation in the hydrodynamic journal bearing. Even with high noisy data, the average relative error related to flowrate identification is around 4% and the maximum relative error is about 10%, observed only at high oil supply flowrates. For low oil supply flowrates, the oil starvation identification is significantly accurate, being the more critical cases in the operation of rotating machinery.

In addition, the EKF was able to successfully identify the oil supply flowrate drops in real-time occurring independently in the bearings. Thus, the proposed method represents a promising tool for real application on monitoring of the faults evolution in the lube oil system, avoiding the occurrence of critical oil starvation of the hydrodynamic bearings that tends to cause sudden maintenance stops and shutdown on rotating machinery.

## Acknowledgments

The authors would like to acknowledge CNPq for the financial support.